\documentclass[11pt]{article}
\usepackage{amsfonts, amssymb, graphicx, a4, epic, eepic, epsfig,
setspace, lscape}

\def\inst#1{$^{#1}$}

\newtheorem{theorem}{Theorem}[section]
\newtheorem{lemma}[theorem]{Lemma}
\newtheorem{proposition}[theorem]{Proposition}

\newtheorem{definition}[theorem]{Definition}
\newtheorem{corollary}[theorem]{Corollary}
\newtheorem{remark}[theorem]{Remark}

\newcommand{\NN}{\mathbb{N}}

\newcommand{\cG}{{\cal G}}
\newcommand{\cC}{{\cal C}}

\newcommand{\cM}{{\cal M}}
\newcommand{\cK}{{\cal K}}

\newcommand{\cX}{{\cal X}}

\def \a {{\alpha}}
\def \b {{\beta}}
\def \e {{\varepsilon}}

\def \l {{\lambda}}

\def \h {{\eta}}
\def \s {{\sigma}}

\def \g {{\gamma}}
\def \t {{\tau}}
\def \o {{\omega}}

\def \p {{\pi}}

\newcommand{\be}[1]{\begin{equation}\label{#1}}
\newcommand{\ee}{\end{equation}}

\newcommand{\bl}[1]{\begin{lemma}\label{#1}}
\newcommand{\el}{\end{lemma}}

\newcommand{\br}[1]{\begin{remark}\label{#1}}
\newcommand{\er}{\end{remark}}

\newcommand{\bt}[1]{\begin{theorem}\label{#1}}
\newcommand{\et}{\end{theorem}}

\newcommand{\bd}[1]{\begin{definition}\label{#1}}
\newcommand{\ed}{\end{definition}}

\newcommand{\bcl}[1]{\begin{claim}\label{#1}}
\newcommand{\ecl}{\end{claim}}

\newcommand{\bp}[1]{\begin{proposition}\label{#1}}
\newcommand{\ep}{\end{proposition}}

\newcommand{\bc}[1]{\begin{corollary}\label{#1}}
\newcommand{\ec}{\end{corollary}}

\newcommand{\bi}{\begin{itemize}}
\newcommand{\ei}{\end{itemize}}

\newcommand{\ben}{\begin{enumerate}}
\newcommand{\een}{\end{enumerate}}

\begin {document}

\title{Some spin glass ideas applied to the clique problem}

\author{%
Antonio Iovanella\inst{1} \and
Benedetto Scoppola\inst{2}\and
Elisabetta Scoppola\inst{3}}
\date{}

\maketitle

\begin{center}
{\footnotesize
\inst{1} Dipartimento di Ingegneria dell'Impresa, University of Rome ``Tor
Vergata''\\
Via del Politecnico, 1 - 00133 Rome, Italy.\\
\texttt{iovanella@disp.uniroma2.it}\\
\vspace{0.3cm} \inst{2} Dipartimento di Matematica, University of Rome
``Tor Vergata''\\
Via della Ricerca Scientifica - 00133 Rome, Italy\\
\texttt{scoppola@mat.uniroma2.it}\\
\vspace{0.3cm} \inst{2} Dipartimento di Matematica, University of Rome
``Roma Tre''\\
Largo San Murialdo, 1 - 00146 Rome, Italy\\
\texttt{scoppola@mat.uniroma3.it}\\ }
\end{center}

\begin{flushright}
\begin{footnotesize}
{\em Be careful men\\
Search every cook and nanny\\
Uh, hook and granny\\
Uh, crooked fan...\\
uh, search everywhere!\\
}
Doc (Snow White and the seven dwarfs,\\
Walt Disney, 1937)
\end{footnotesize}
\end{flushright}

\begin{abstract}

In this paper we introduce a new algorithm to study some NP-complete
problems. This algorithm is a Markov Chain Monte Carlo (MCMC)
inspired by the cavity method developed in the study of spin glass.
We will focus on the maximum clique problem and we will compare this
new algorithm with several standard algorithms on some DIMACS
benchmark graphs and on random graphs. The performances of the new
algorithm are quite surprising. Our effort in this paper is to be
clear as well to those readers who are not in the field.
\end{abstract}

\eject

\tableofcontents

\section{Introduction}

In the last years Mezard, Parisi, Zecchina~\cite{MPZ},~\cite{MZ}
introduced a class of optimization  algorithms to deal with
K-satisfiability problems. Their strategy was based on the cavity
method introduced in spin glass theory a long time ago  and in
particular on its zero-temperature version, more recently developed
in~\cite{MP}. An important ingredient in their approach seems to be
the locally tree-like structure of the interaction graph.

In the case of the clique problem,  i.e., the study of the maximal
complete subgraph of a given graph $G$, we expect to be very far
from a tree-like structure of the interaction graph even locally,
for instance when $G$ is a random graph. We introduce in this paper
a new algorithm to treat this problem, based again on the cavity
method but in a completely different way.

This algorithm represents a first step in the application of the
cavity idea that will be developed in a forthcoming paper. On the
other hand this algorithm is sufficiently simple so that its
behavior can be studied at least on random graphs providing some
explanation of the difficulty of the problem. The algorithm
introduced in this paper represents an heuristical search of cliques
in the sense that the optimality of the result is not guaranteed.
For a recent review on the numerical approach to the clique problem
see e.g.~\cite{CF} and references therein.

\subsection{Definitions}

Let $G=(V,E)$ be a graph. A graph $g$ is  a {\it subgraph}  of $G$,
$g\subseteq G$, if its vertex set  $V(g)\subseteq V$ and its edges
$E(g)\subseteq E$. For any $A\subset V$ we denote by $G[A]$ the
{\it graph induced by } $A$ in $G$:

\be{Ga}
G[A]=(A,E(G[A])), \quad \hbox{ with }\quad E(G[A]):=\{(i,j)\in E:\,
i,j\in A\}
\ee

We will denote by $\cK(G)$ the set of {\it complete subgraphs or
cliques} of $G$ and by $\cM ax\cC l(G)$ the set of {\it maximum
cliques} in $G$:

\be{maxcl}
\cM ax\cC l(G):=\{ g\in\cK(G):\; |V(g)|=\max_{g'\in\cK(G)}|V(g')|\}
\ee
where $|B|$ denotes the cardinality of the set $B$.

We call {\it clique number} of the graph $G$, $\o(G)$, the
cardinality of the vertex set of any maximum clique in $G$, i.e.,
$\o(G)=|V(g)|$ with $g\in \cM ax\cC l(G)$.

There are several versions of the problem of the determination of
the clique number and of the maximum clique set of a given graph
$G$. We recall here the most cited form.
\bi
\item[] {\bf Clique problem:} given a graph $G=(V,E)$ and a positive
integer
 $k\le |V|$, does $G$ contain a complete subgraph of size $k$ or more?
That is, does $\o(G)\ge k$ hold? \ei As it is well known
(see~\cite{GJ}) the clique problem is a NP-complete problem. There
are other  famous NP-complete  problems equivalent to the clique
problem as the vertex covering and the independent set, defined as
follows: \bi
\item[]{\bf Vertex covering:}  given a graph $G=(V,E)$ and a positive
integer  $k\le |V|$, is there a vertex cover of size $k$ or less for
$G$, i.e., a subset $V'\subseteq V$ with $|V'|\le k$ such that for
each edge  $(u,v)\in E$ at least one of $u$ and $v$ belongs to $V'$?
\item[]{\bf Independent set:}  given a graph $G=(V,E)$ and a positive
integer  $k\le |V|$, does $G$ contain an independent set of size $k$
or more, i.e., a subset $V'\subseteq V$ such that $|V'|\ge k$ and
such that no two vertices in $V'$ are joined by an edge in $E$? \ei
The equivalence of these problems is proved for instance
in~\cite{GJ} lemma 3.1 pg.54.

\subsection{The case of random graphs}\label{randgraph}
Consider the set $\cG(n,d)$ of random graphs with fixed density $d$,
i.e. of graphs $G(V,E)$ having as vertex set $V=\{1,2,...,n\}$ and
in which the edges are chosen independently with probability $d$.

To study 
the size of the largest clique of a graph $G(V,E)\in \cG(n,d)$ 
one can argue as follows. Let $Y_r$ be the number of complete subgraph with $r$
vertices in a graph $G(V,E)\in \cG(n,d)$. It is immediate to show
that \be{yr} E(Y_r)={n\choose r}d^{r\choose 2} \ee Let us consider
the value $r_0(n)$ of $r$ such that $E(Y_r)=1$. Writing (\ref{yr})
in terms of Stirling approximation and denoting $b=1/d$ we have that
such value $r_0(n)$ is given by \be{r0}\label{ro}
r_0(n)=2\log_bn-2\log_b\log_bn+2\log_b(e/2)+1+o(1) \ee

The clique number $\o(G)$ of a graph $G(V,E)\in \cG(n,d)$ tends, for
$n\to\infty$, to be very near to $r_0(n)$. More precisely, it is
possible to prove the following result (see \cite{bollobas}): for
almost all the graphs $G\in \cG(\mathbb{N},d)$ there is a constant
$m_0(G)$ such that for all $n\ge m_0(G)$ and for almost all $G_n$
subgraph of $G$ with vertex set $|V|=n$ \be{omega1}
\vert\o(G_n)-2\log_bn+2\log_b\log_bn-2\log_b(e/2)-1\vert\le{3\over
2} \ee

Despite the fact that the asymptotic value of $\o(G_n)$ has such a
small variability, it is well known 
that the large
cliques of a random graph are very difficult to find. This is due to
the fact that the expression of $E(Y_r)$, which has its maximum for
an $r$ that is roughly $r_0(n)/2$, decreases very rapidly when $r>
r_0(n)/2$. Hence, while it is easy (e.g. with a greedy algorithm) to
find cliques whose size is of the order of $\log_bn$, the
probability that one of such cliques is a subset of a clique with
the  size $(1+\e)\log_bn$ is of the order of $n^{-\a(\e)\log n}$ for
all $\e>0$, and hence is more than polynomially small (see also~ \cite{jerrum}). 

This difficulty in finding large clique of random graphs has a
numerical evidence even for $n$ quite small, as it will be shown
later.

\subsection{The statistical mechanics approach}

We recall here very briefly the main ideas of the statistical
mechanics approach to combinatorial optimization problems.

The cost function of the optimization problem (OP) can be view as
the energy function $H(x)$, usually called {\it Hamiltonian}, of a
statistical mechanics (SM) model where instances of the OP are
considered as configurations $x\in\cX$ of the SM model. The optimal
configurations correspond to the ground states in the SM language.
(See for instance~\cite{MPV}). Ground states in SM are the
configurations where the Gibbs measure $\p(x)={1\over Z}e^{-\b
H(x)}$ is concentrated in the limit of zero temperature
($\b\to\infty$, being $\b$ the inverse temperature); the
normalization constant $Z$ is usually called {\it partition
function}. This means that to determine  the ground states is
sufficient to perform  a random sampling at low temperature. To this
purpose we can apply the Monte Carlo method. The main idea of this
method is to define a {\it Markov chain Monte Carlo} (MCMC) on the configuration
space $\cX$, with transition probabilities $P(x,x')$ such that the
transition probability in $n$ steps, $P^n(x,x')$, of the chain
converges to $\p(x')$ as $n\to\infty$. This convergence is due to
the ergodic theorem if for instance the transition probabilities
satisfy a {\it detailed balance condition} w.r.t. the Gibbs measure
$\p$: \be{dbc} \p(x) P(x,x')=\p(x') P(x',x) \ee The strategy of the
MCMC method is then the following \bi
\item[-] start from a configuration $x_0$
\item[-] look at the random evolution of the chain starting
from it, $x_0,x_1,...x_n$, for a ``sufficiently long time'' $n$
\item[-] for the final state $x_n$ we have $P(x_n=x)\sim \p(x)$.
\ei

The main difficulty in applying this procedure is due to metastable
states. Indeed local minima of the energy $H(x)$ can capture the
evolution $x_t$ of the chain for very large time intervals if the
temperature is low. So the main problem in applying MCMC method is
to define what ``sufficiently long time'' means.

A strategy to escape the problem of metastable states is to change
the temperature during the evolution of the chain. This is known as
{\it simulated annealing}. Since for high temperature the process
leaves local minima much easily, one can look at a suitable
annealing in order to avoid to remain captured in metastable states.
See for instance~\cite{KGV} for the use of simulated annealing in
optimization problems.

>From a rigorous point of view  the main point in applying the MCMC
method is to estimate the {\it mixing time} of the chain, that is
the time $n$ necessary to have that $P(x_n=x)$ and  $ \p(x)$ are
sufficiently close each other, uniformly in $x_0$. (See for
instance~\cite{mixing} for precise definitions.)

As an example for the clique problem on a graph $G\in\cG(n,1/2) $ we
can consider as in~\cite{jerrum} the following MCMC. The state space
$\cX$ of the chain is the collection of all cliques in G. To each
clique $x\in\cX$ we associate a weight $w(x)=\l^{|x|}$ where $|x|$
denotes the number of vertices of $x$ and $\l\ge 1$ is a real
parameter. We can describe this  weight in terms of a Gibbs measure
$\p(x)={w(x)\over Z} $ with $H(x)=-|x|$ and $\l=e^{\b}$. The
transition probability $P(x,x')$ is different from zero only if the
cliques $x$ and $x'$ have a symmetric difference (as sets of
vertices) less or equal to one. In this case if $x'\supset x$ we put
$P(x,x')={1\over n}$ and if  $x'\subset x$ we put $P(x,x')={1\over
\l n}$. The probability $P(x,x)$ is obtained by normalization. It is
immediate to verify that these transition probabilities satisfy the
detailed balance condition (\ref{dbc}).

For this dynamics Jerrum proves that there exists an initial state
from which the expected time to reach a clique of size at least
$(1+\e)\log_2n$ is super-polynomial in $n$. The crucial point in
this proof is to show that there are few cliques that can grow up to
this size  $(1+\e)\log_2n$. More precisely a clique of size $k$ is
called {\it m-gateway}  if there exists a path of the chain going
from this clique to a clique of size $m$ through cliques of size at
least $k$, then it is proved in~\cite{jerrum} that the density of
$m$-gateways in the set of $k$-cliques is super-polynomially small
for $k=\lceil (1+{2\over 3}\e)\log_2n\rceil$ and
$m=\lceil(1+\e)\log_2n\rceil$. Due to the fact that $m$-gateways
have to be visited in reaching cliques larger or equal to $m$, then
these $m$-gateways represent a bottleneck for the dynamics and their
low density can be used to prove that the mixing time is
super-polynomial in $n$.

\section{A Hamiltonian for the clique problem, the vertex covering
and the independent set}

We consider the  space $\cX:= \{0,1\}^V$ of lattice gas
configurations on $V$; on the configuration space (or state space)
$\cX$ we define an Ising Hamiltonian with an antiferromagnetic
interaction between non-neighbor sites:

\be{defham2}
H(\s):=   \sum_{(i,j)}J_{ij}\s_i\s_j  -{h}\sum_{i\in V}\s_i
\ee
where
\begin{equation}
J_{ij}=\cases{0&if $(i,j)\in E$\cr
1&if $(i,j)\in E^c$\cr}\label{Jij}
\end{equation}
with $E^c:=\{(i,j)\not\in E; \, i,j\in V\}$ and $h>0$.

It is easy to prove that if $h<1$ then the minimal value of  $H(\s)$
is obtained on configurations with support on the vertices of a
maximum clique. First of all we prove that $H(\s)$ is minimal on
configurations $\s$ such that  $G(\s)\in\cK(G)$. We denote with the
same letter a configuration and its support; for instance when we
write $i\in\s$ we mean a site $i$ in the support of $\s$. Indeed for
every $\s$ such that $G(\s)\not\in\cK(G)$ $\s=C\cup A$ with $G(C) $
a maximum clique in $G(\s)$ and $|A|\ge 1 $, then for any  $i\in A$
we have $H(\s)>H(\s\backslash i)$. This is due to the fact that
\be{01} H(\s)=H(\s\backslash i)+\sum_jJ_{ij}\s_j  -h \ee and if
$G(C) $ is a maximum clique in $G(\s)$  then $\sum_jJ_{ij}\s_j \ge
\sum_{j\in C}J_{ij} \ge 1$. As a second step we note that if $\s$ is
such that  $G(\s)\in\cK(G)$ then $H(\s)=-{h\over 2}|\s|$, so that we
can immediately conclude that $H$ is minimal on the maximum cliques.

If we consider the opposite interaction:

\begin{equation}
\bar J_{ij}=\cases{0&if $(i,j)\in E^c$\cr
1&if $(i,j)\in E$\cr}\label{Jij1}
\end{equation}
then the same Hamiltonian (\ref{defham2}) with interaction $
\bar J$ is minimal on configurations with
zeros on a minimal vertex cover and ones on the maximum independet set.

In the case of a random graph $G$, i.e., when the interaction
variables $J_{ij}$ are i.i.d.r.v., the Hamiltonian (\ref{defham2})
is similar to the Hamiltonian of the SK model. The main differences
are that our configurations are in lattice gas variables instead of
spin variables and the interaction variables have no zero mean.
Instead of a symmetry property we have now a control on the sign of
the interaction term of the Hamiltonian.

\section{Some algorithms for the clique problem}\label{cliquealgo}

In this section we define three different algorithms for the clique
problem that will be used for the numerical comparison developed in
the final section. The first and the second are ``standard''
algorithms; the third algorithm is a MCMC defined by means of the
Hamiltonian (\ref{defham2}).

\subsection{A greedy algorithm, $\cal{G}$}

The first algorithm we introduce is a fast and  greedy heuristic,
denoted from now on by $\cal{G}$. The underlying idea is to start
from a configuration $\s$ with $\s_i =0, \forall i \in V$ and then
select at random a vertex $j$, set $\s_j = 1$ and then delete all
its non adjacent vertices. In the next step another vertex is
selected at random among the remaining vertices and again all its
non-adjacent vertices are deleted. The process stops when it is not
possible to select other vertices, i.e., a maximal complete subgraph
is found, i.e., a clique not strictly contained in other cliques.

\subsection{A dismantling algorithm, $\cal{D}$ }

The second algorithm, denoted  by $\cal{D}$ in the following, is
another fast heuristic. It starts with an initial configuration $\s$
that has ones everywhere. The algorithm considers, at each step, the
degree of each vertex $i$ with $\s_i =1$ and selects the one (say
$j$) with the smallest degree. Then it sets $\s_j = 0$ and decreases
by one unit the degree of all its adjacent nodes in the graph and
repeats the procedure until the minimum value of all the degrees is
$k-1$ where $k$ is the number of sites in $\s$ with $\s_i = 1$,
i.e., the sites of a clique of cardinality $k$.
Note that in principle the resulting clique could be not maximal.

The rationale of this algorithm is to start from the whole graph
and then, at each step, dismantle it vertex by vertex until a clique
is found.

\subsection{A Monte Carlo algorithm $\cal{MC}$}

We can apply the ideas developed in section 1.3 to the Hamiltonian
defined in section 2  for  the clique problem. For clarity we
consider the Metropolis choice: for $\s' \ne \s$ we take

\be{metropolis}
P(\s',\s)=q(\s',\s) e^{-\b[H(\s)-H(\s')]_+},
\ee

where $[\cdot]_+$ denotes the positive part and $q(\s',\s)$ is a
symmetric, positive connectivity  matrix independent of $\b$ with
$q(\s',\s)>0$ only if $\s$ and $\s'$ are different in a single site.

We note that in the limit $\b\to\infty$ and $h\in (0,1)$ fixed,
starting from the configuration which is zero everywhere, this
algorithm is equivalent to the greedy algorithm since $P(\s,\s')=0$
if $H(\s')>H(\s)$. Thus $\{\s(t)\}_{t\in\NN}$ is a growing sequence
of complete graphs. In the case $\b\to\infty$ but $h\to 0$ as
${1\over\b}$ we see that this Monte Carlo algorithm is equivalent to
the Jerrum algorithm on cliques recalled as an example in section
1.3.

\section{A new algorithm inspired by the cavity method, ${\cC}$}

In this section we introduce a new algorithm to find maximum cliques
of a  graph. The key idea is inspired by the notion of {\it cavity
field} introduced in statistical mechanics to analyze the ground
states, that is configurations minimizing the energy (\cite{MPV}).
The cavity method at zero temperature is described in detail
in~\cite{MP} in the case of a spin glass on a lattice with a local
tree like structure. This method is equivalent to the replica method
and can be used at different levels of approximation corresponding
to the replica symmetric solution and to the one step replica
symmetry breaking level. The main idea is to compute in the limit of
infinite number of spins the value of the energy density of the
ground state by  an iterative procedure. Indeed one can study the
effect of the addiction of a spin or of a bond to the system looking
for equations for the corresponding average energy shift.

We do not use the cavity method in our algorithm but we use the idea
that if you select  a spin the effect of the other spins can be
described in terms of a local field, that we will call cavity field,
as in the case of the cavity method.

More precisely, consider the Hamiltonian defined in (\ref{defham2})
and consider the canonical ensemble, i.e., the set of configurations
$\s\in\cX$ such that $\sum_{i\in V}\s_i=k$. Up to a constant we have
that $H(\s)=\sum_{(i,j)} J_{ij}\s_i\s_j$. If for each $i\in V$ we
define the cavity field: \be{hi} h_i(\s)=\sum_{j\not=
i}J_{ij}\s_j+h(1-\s_i) \ee we have immediately that
$H(\s)=\sum_{i\in V}h_i(\s)\s_i$. For a given choice of the fields
$\{h_i\}_{i\in V}$ the minimal energy is clearly obtained on the
configurations with support on the sites corresponding to the $k$
minimal values of $h_i$. But here the cavity fields depend on the
configuration itself and then it is more difficult to determine the
ground states. To this purpose we  introduce a new Hamiltonian:
\begin{equation}
H(\sigma,\sigma',h,k)=\sum_{i\neq
j}J_{ij}\sigma_i\sigma'_j\label{H}+h(k-\sum_i\sigma_i\sigma'_i)
\end{equation}
defined on pairs of configurations $\sigma,\sigma'$ such that
$\sum_i\sigma_i=\sum_i\sigma'_i = k$, with $k\in\NN$ and $h>0$. The
hamiltonian can be rewritten in terms of the interaction of the
configuration $\sigma'$ with each site $i$ (cavity field $h_i$) in
the following way
\begin{equation}
H(\sigma,\sigma',h,k)=\sum_{i\neq
j}J_{ij}\sigma_i\sigma'_j+h(k-\sum_i\sigma_i\sigma'_i)
=\sum_ih_i(\sigma)\sigma'_i\label{Hbar}
\end{equation}
with $h_i=h_i(\sigma)$ defined in (\ref{hi}). Hence the cavity field
$h_i$ in the site $i$ represents the number of sites $j$ with
$\sigma_j=1$ that are not nearest neighbors of the site $i$ plus a
contribution $h$ that is present when the configuration $\sigma$ is
not supported on the site $i$. Note that $H(\sigma,\sigma,h,k)$
corresponds to the Hamiltonian (\ref{defham2}) in the framework of
the canonical ensemble corresponding to $k$. We also want to stress
that this new Hamiltonian is non-negative and if $k\le\o(G)$ its
value is zero (so minimal) only on pairs of configurations $\s,\s'$
such that $\s=\s'$ with support on a clique with $k$ vertices.

The idea of the algorithm is the following: start from a random
configuration $\sigma$ with fixed $k$, and choose a new
configuration $\sigma'$ picking randomly $k$ sites, each site having
a relative weight $w_i=e^{-\beta h_i(\sigma)}$ for some $\beta>0$,
and define for this sites $\sigma'_i=1$, while for the others
$\sigma'_i=0$. Then repeat this procedure iteratively. After each
iteration compute the quantity $ H(\sigma,\sigma',h,k)$. This
dynamics defines a MCMC on $\cX$ that satisfies the detailed balance
condition with respect to the stationary measure
\begin{equation}
\Pi_\sigma=\frac
{\sum_\tau e^{-\beta H(\sigma,\tau,h,k)}}
{\sum_{\tau,\sigma} e^{-\beta H(\sigma,\tau,h,k)}}
\label{Pi}
\end{equation}
Indeed since each vertex $j$ is chosen to have
$\sigma'_j=1$ with weight $w_i$, the transition probability of the
process $P(\sigma,\sigma')$ has the following form
\begin{equation}
P(\sigma,\sigma')=\frac{e^{-\beta  H(\sigma,\sigma',h,k)}}
{\sum_\tau e^{-\beta H(\sigma,\tau,h,k)}}\label{pss'}
\end{equation}
Due to the symmetry of the couplings $J_{ij}=J_{ji}$ we have
\begin{equation}
\Pi_\sigma P(\sigma,\sigma')=
\frac
{\sum_\tau e^{-\beta H(\sigma,\tau,h,k)}}
{\sum_{\tau,\sigma} e^{-\beta H(\sigma,\tau,h,k)}}
\frac{e^{-\beta  H(\sigma,\sigma',h,k)}}
{\sum_\tau e^{-\beta H(\sigma,\tau,h,k)}}=
\end{equation}
\begin{equation}
=\frac
{\sum_\tau e^{-\beta H(\sigma',\tau,h,k)}}
{\sum_{\tau,\sigma'} e^{-\beta H(\sigma',\tau,h,k)}}
\frac{e^{-\beta H(\sigma',\sigma,h,k)}}
{\sum_\tau e^{-\beta H(\sigma',\tau,h,k)}}= \Pi_{\sigma'}
P(\sigma',\sigma)
\label{bildett}
\end{equation}
and therefore $\Pi_\sigma$ is the unique stationary measure of our
process.

Note that if the parameter $\beta$ is very large and $k\le\o(G)$,
the stationary measure is concentrated exponentially in $\beta$ on
the $\sigma$'s such that there exists a clique supported by the
configuration $\sigma$: actually if the support of $\s$ is not a
clique $ H(\sigma,\tau,h,k)>0$ for all configurations $\tau$ and the
probability of the configuration $\sigma$ is exponentially small.

\subsection{Implementation of the algorithm ${\cC}$ and some remarks
on its mixing time}

To realize a single step of the Markov chain with transition
probabilities defined in (\ref{pss'}) we proceed as follows. \bi
\item[1.] Starting from a configuration $\s$, compute the cavity field
$h_i(\s)$ for each vertex $i$.
\item[2.] To sample the new configuration $\s'$ with probability
 (\ref{pss'}) we perform a Kawasaki-like algorithm
 $\h(0),\,\h(1),\,...\h(s),\,...$, starting at $\h(0)=\s$.
At each step $s$ this Kawasaki procedure is the following: pick randomly
a couple of vertices $(i,j)$ such that $\h_i(s)=1$ and $\h_j(s)=0$
and define $\h(s)^{(i,j)}$, the configuration obtained by $\h(s)$ by
exchanging the occupation variables in the sites $i$ and $j$. Then
$\h(s+1)= \h(s)^{(i,j)}$ with probability $e^{-\b[h_j(\s)-h_i(\s)]_+}$.
Since $H(\s,\h(s))=\sum_{i\in V}h_i(\s)\h(s)_i$ we have
 $H(\s,\h(s+1))- H(\s,\h(s))= h_j(\s)-h_i(\s)$ so that the invariant
measure of this Kawasaki chain is \be{PK}
\Pi^K_{\s'}=\frac{e^{-\beta  H(\sigma,\sigma',h,k)}} {\sum_\tau
e^{-\beta H(\sigma,\tau,h,k)}} \ee as requested in (\ref{pss'}). \ei

Since this measure $\Pi^K_{\s'} $ is a product measure we note that
step 2, i.e., the Kawasaki procedure, quickly reaches its
equilibrium, in a time of order $nk$. Much more complicated is an
estimate for the mixing time of the chain ${\cC}$. Here we can only
make some initial remarks on this problem.

First of all we note that the function $H(\s(t),\s(t+1),h,k)$ is a
non-increasing function of $t$ in the limit $\b\to\infty$ along a
typical path $\{\s(t)\}_t$ of the chain $\cC$. Indeed in the limit
of zero temperature, the configuration $\s(t+1)$ minimizes the
Hamiltonian

\be{minH}
\min_{\s}H(\s(t),\s,h,k)=H(\s(t),\s(t+1),h,k)=H(\s(t+1),\s(t),h,k)
\ee and  $\s(t+2)$ is such that \be{minH1}
\min_{\s}H(\s(t+1),\s,h,k)=H(\s(t+1),\s(t+2),h,k)\le
H(\s(t+1),\s(t),h,k) \ee So the {\it trap configurations} for the
dynamics $\cC$ at zero temperature are the configurations $\s$ such
that $\min_{\t}H(\s,\t,h,k)=H(\s,\s,h,k)$. The cavity fields
$h_i(\s)$ have values in the set
$\{q+rh\}_{q\in\{0,1,...,k\},\,r\in\{0,1\}}$ so we can define the
different levels of the cavity fields of $\s$, i.e., for each
$q\in\{0,1,...,k\}$ and $r\in\{0,1\} $ we define $I_{q,r}:=\{i\in
V:\, h_i(\s)=q+rh\}$. The configurations $\t$  minimizing
$H(\s,\t,h,k) $ have support on the sites belonging to the lowest
levels of the cavity fields $h(\s)$. This means that $\s$ is a trap
if $h_{max}(\s):=\max_{i\in\s}h_i(\s)< h_j(\s)$ for each
$j\not\in\s$. On the other hand, in the case of random graphs, we
know the distribution of the cavity field in sites $j\not\in\s$.
Indeed for these sites we have $h_j(\s)= ML_j(\s)+h$ where
$ML_j(\s)$ denotes the number of missing links from $j$ to the set
$\s$ (the support of $\s$). Due to the fact that $ML_j(\s)$ and
$ML_{j'}(\s)$ are independent variables for $j,j'\not\in\s$ with a
binomial distribution, we also know the distribution of the numbers
$|I_{q,1}|$ of sites $j\not\in\s$ with cavity field $h_j(\s)= q+h$:
\be{dI} P(|I_{q,1}|=l)= {n-k\choose l}p^l (1-p)^{n-k-l} \ee where
\be{PML} p\equiv p(q,k):= P(ML_j(\s)=q)= {k\choose q} 2^{-k} \ee in
the case of random graph with density ${1\over 2}$. The quantity
\be{gap} G(\s):=\min_{j\not\in\s}h_j(\s)- h_{max}(\s) \ee can be
called the {\it gap of the trap}.

If $k=(1+\e)\log_2n$ with $ \e>0$ and if $ q\ll k$
we have that for large $n$
\be{d1I}
P(|I_{q,1}|=0)=  (1-p)^{n-k}\sim 1-n^{-\e}
\ee
 so with large probability the lowest levels corresponding to
$r=1$, i.e., to sites not in $\s$, are empty.

We notice that in order to really leave a trap, we have to change
enough many sites in a single step of the dynamics. Small changes
produce configurations immediately coming back to the trap. Indeed
starting from a trap $\s$ with gap $\g$, denote by $\s'$  the
configuration obtained in a single step of the dynamics and by $l$
the number of changed sites, i.e., $l=|\{ i;\, \s_i\not=\s'_i\}|$.
We have that $|h_i(\s)-h_i(\s')|\le l+h$ for each site $i$. So if
$l<{\g\over 2}-h$ we have that the new cavity field $h(\s')$ has the
lowest levels again containing the sites of $\s$ and so with large
probability the dynamics in the following steps will came back in
$\s$.

Again we can apply the Jerrum argument. If $\s$ is almost a clique
--i.e., $h_{max}(\s)$ is small but the maximal clique contained in
$s$, say $\s_0$, is of size $ k_0=(1+{2\over 3}\e)\log_2n $ and
$\s_0$  is not a $k$-gateway-- with probability near to one we have
a gap of $\s$ of order $ak$ with $a<1$ but strictly positive. This
means that to escape the trap we have an energy barrier that is a
positive fraction of $k^2$ since, if $\s$ is not a $k$-gateway,  a
number of sites proportional to $k$ has to be changed in the
non-empty levels of type  $I_{q,1} $.

For this reason we expect that our algorithm has a non-polynomial
mixing time of order $n^{a\log n}$. However in the following section
we will show that this non-polynomial mixing time becomes evident
only when $n$ is very large. On DIMACS random graphs we get better
results than the other algorithms.

Moreover we can gain from our analysis of traps a more precise
knowledge of the energy landscape, suggesting improvements of our
algorithm. This is the subject of a further paper where the
numerical aspects of the problem will be discussed in more details.

\section{Numerical comparison}

In this section we briefly give some numerical results on the
algorithm introduced in the previous section. In particular we will
compare our algorithm with the ``standard'' ones recalled in
Section~\ref{cliquealgo} on two groups of graphs: DIMACS benchmark
graphs~\cite{johnson} and random graphs. A more complete numerical
analysis will be given in~\cite{ISS}.

\subsection{Experimental details}

All our algorithms, the greedy $\cal{G}$, the dismantling $\cal{D}$,
the Monte Carlo $\cal{MC}$ and Cavity $\cal{C}$ are implemented in C
language and performed on a $2.5$GHz Power Mac G5 Quad processors
machine with Mac OS X v10.4 Tiger and 8Gb of RAM and compiled with
gcc and considering the -O2 switch. As required by the rules of the
Second DIMACS Implementation Challenge~\cite{johnson}, we provide in
Table~\ref{machineDIMACS} the user times in seconds performed by one
processor on our computer.

\begin{table}
\begin{center}
\begin{tabular}{ccccc}\hline
{\tt r100.5}&{\tt r200.5}&{\tt r300.5}&{\tt r400.5}&{\tt r500.5}\\
\hline
 0.01  & 0.09  & 0.77  & 4.47 & 16.83  \\
\hline
\end{tabular}
\caption{User times for DIMACS machine benchmarks instances.}\label{machineDIMACS}
\end{center}
\end{table}

\subsection{Numerical results on DIMACS benchmark graphs}

The Center for Discrete Mathematics and Theoretical Computer Science
(DIMACS) makes available on its web site ({\tt
ftp://dimacs.rutgers.edu/pub/challenge/graph/benchmarks}) a suite of
$79$ benchmark graphs for the maximum clique size problem. Such
benchmarks constitute an important base point in order to evaluate
the performances of new algorithms in this topic. They were
generated by means of different criterions and the set includes:

\bi
 \item Random graph ({\tt C{\it n}.{\it d}} and {\tt DSJC{\it n}.{\it d}}, being $n$ the size and $d$ the density);
 \item Steiner triple graph ({\tt MANN{\it n}});
 \item Brockington graph ({\tt brock{\it n}\_{\it y}}, with parameter $y = 1$, $2$, $3$, $4$);
 \item Sanchis graph ({\tt gen{\it n}\_p0.9\_{\it x}}, {\tt san{\it n}\_0.{\it y}\_{\it z}}, {\tt sanr{\it n}\_0.{\it y}}, with parameters $x = 44$, $55$, $65$, $75$, $y = 5$, $7$, $9$ and $z =  1$, $2$, $3$);
 \item Hamming graph ({\tt hamming{\it x}-{\it y}} with parameters $x = 6$, $8$, $10$ and $y = 2$, $4$);
 \item Keller graph ({\tt keller{\it x}}, with parameter $x = 4$, $5$, $6$);
 \item P-hat graph ({\tt p-hat{\it n}-{\it x}}, with parameter $x= 1$, $2$, $3$);
 \item Pardalos graph ({\tt c-fat{\it n}-{\it x}}, with parameter $x = 1$, $2$, $5$, $10$);
\ei

For additional details and references the reader could
see~\cite{johnson}.

In Table~\ref{tabella1} we report the results for a selection of the
$37$ instances belonging to the Second DIMACS Implementation
Challenge and in Tables~\ref{tabella2} a selection of the remaining
$42$. The tables are organized as follows. The first column report
the name of the instance, the following three columns its
characteristics, i.e., number of nodes $n$, number of arcs $m$ and
density $d$. Successively, we report the results for the cavity
algorithm in terms of best achieved value and CPU time. The
remaining columns are related to the values achieved by the
algorithm presented in Section~\ref{cliquealgo}. In particular,
column $\cal{G}$ reports the best value achieved on 100 run of the
greedy algorithm and columns $\cal{D}$ and $\cal{MC}$ report both
best achieved clique and CPU time, respectively. Note that for the
sake of simplicity we do not report the CPU time for $\cal{G}$
because it was always equal to $0.000$.

Let us close this section with some finale remarks on the
performances of the $\cal{D}$ and $\cal{C}$ algorithms. First of all
we want to stress that despite to its simplicity, $\cal{D}$ performs
quite well on many instances, especially when the density is high
and exact results are difficult to obtain. As far as $\cal{C}$ 
is concerned we
consider its performances quite promising.

\subsection{Numerical results on random graphs}

In order to give a deeper analysis of the performance of our
algorithm on random graphs, our experiments were extended to a
collection of big instances built by mean of a random graph
generator. In fact, even thought the DIMACS collection includes some
random instances, the number of nodes are no greater then $4000$.
For this reason, we implemented a random graph generator able to
build instances with a fixed number of nodes and density limited
only by the space occupancy of the graph on the physical memory
existing on the computer. Our choice was to build a collection of
fifteen instances with $n =2^i$ for $i=7,8,...,14$ i.e., for
$n\in\{128, 256, 512, 1024, 2048, 4096, 8192, 16384\}$ and density
$d=\{0.5, 0.9\}$. The name of the instances considers first the
prefix {\tt tbb}, then the number of nodes, the density and finally
the extension {\tt clq.b}. Again, note that the instances follows
the rules provided by the DIMACS.

These instances are available for further research on the web on the
home page of one of the
co-author\footnote{http://www.disp.uniroma2.it/Users/iovanella/clique}.

On the smaller graphs we obtained the certified values of the clique
number by using the program Cliquer. This is a branch and bound
algorithm given in~\cite{NO}. As it is clear from the Table~\ref{tabella5} the
computational times of Cliquer are too long to apply it to the
larger instances.

In Tables~\ref{tabella3} and~\ref{tabella4} are reported the results
on our instances, for $d = 0.5$ and $d = 0.9$ respectively.

{\bf Acknowledgments:}

First of all we want to thank A.Sinclair who introduced us to the
subject providing useful suggestions and references. We are in debt
to G.Parisi for many fruitful discussions; in particular, he first
proposed to look at the canonical ensemble, which constituted one of
the main ingredients of our approach. We also benefitted from our
many lunch breaks with F.Martinelli who represented a continuous
rich source of salient remarks. We want also to thank R.D'Autilia
and F. Zamponi for useful discussions and P.Dell'Olmo for his warm 
encouragement.
Thanks also to D.Brydges.

\begin{landscape}
\begin{table}
\begin{small}
\begin{center}
\begin{tabular}{l|cccc|cc|c|cc|cc}
\hline
&&&&&\multicolumn{2}{|c|}{$\cal{C}$} & $\cal{G}$ & \multicolumn{2}{|c|}{$\cal{D}$}& \multicolumn{2}{|c}{$\cal{MC}$}\\
\hline
DIMACS benchmarks         & $n$  &   $m$   &  $d$  & $   \o(G)$ &      $k$ & Time ($s$) &   $k$ &          $k$ &  Time($s$)          &          $k$ &  Time($s$)    \\
\hline\hline
\texttt{C125.9}           &  125 &    6963 & 0.898 &         34 & {\bf 34} &   0.060 & 23 &   32 &  0.000 &       28 &    0.640 \\
\texttt{C250.9}           &  250 &   27984 & 0.899 &         44 & {\bf 44} &   0.270 & 29 &   39 &  0.000 &       35 &    2.660 \\
\texttt{C500.9}           &  500 &  112332 & 0.900 &         57 & {\bf 57} &   2.670 & 36 &   47 &  0.000 &       41 &   15.340 \\
\texttt{C1000.9}          & 1000 &  450079 & 0.901 &         68 & {\bf 68} &  20.970 & 43 &   53 &  0.080 &       46 &   77.840 \\
\texttt{C2000.9}          & 2000 & 1799532 & 0.900 &   $\geq$80 &       77 &  75.760 & 49 &   56 &  0.330 &       52 &  371.460 \\
\texttt{DSJC500.5}        &  500 &   62624 & 0.502 &         14 &       13 &   1.290 &  9 &    8 &  0.010 &       11 &   18.770 \\
\texttt{DSJC1000.5}       & 1000 &  499652 & 0.500 &         15 & {\bf 15} &  15.580 & 10 &   10 &  0.070 &       12 &   95.970 \\
\texttt{C2000.5}          & 2000 &  999836 & 0.500 &   $\geq$16 & {\bf 16} &   9.900 & 12 &    9 &  0.310 &       13 &  409.530 \\
\texttt{C4000.5}          & 4000 & 4000268 & 0.500 &   $\geq$18 & {\bf 18} & 104.020 & 12 &   12 &  1.380 &       15 & 1889.440 \\
\texttt{MANN\_a27}        &  378 &   70551 & 0.990 &        126 &      124 &   6.600 & 90 &  117 &  0.000 &      110 &    5.380 \\
\texttt{brock200\_2}      &  200 &    9876 & 0.496 &         12 & {\bf 12} &   0.000 &  8 &    8 &  0.000 &       10 &    2.330 \\
\texttt{brock200\_4}      &  200 &   13089 & 0.658 &         17 & {\bf 17} &   0.040 & 11 &   12 &  0.000 &       14 &    2.140 \\
\texttt{brock400\_2}      &  400 &   59786 & 0.749 &         29 &       25 &   1.050 & 17 &   21 &  0.000 &       20 &    9.100 \\
\texttt{brock400\_4}      &  400 &   59765 & 0.749 &         33 &       25 &   1.340 & 17 &   20 &  0.010 &       20 &    9.040 \\
\texttt{brock800\_2}      &  800 &  208166 & 0,651 &   $\geq$21 & {\bf 21} &   0.350 & 14 &    7 &  0.020 &       17 &   33.050 \\
\texttt{brock800\_4}      &  800 &  207643 & 0,650 &   $\geq$21 & {\bf 21} &   0.610 & 14 &   13 &  0.040 &       17 &  333.070 \\
\texttt{gen200\_p0.9\_44} &  200 &   17910 & 0.900 &         44 & {\bf 44} &   0.360 & 27 &   31 &  0.000 &       33 &    1.600 \\
\texttt{gen200\_p0.9\_55} &  200 &   17910 & 0.900 &         55 & {\bf 55} &   0.030 & 28 &   35 &  0.000 &       41 &    1.600 \\
\texttt{gen400\_p0.9\_55} &  400 &   71820 & 0.900 &   $\geq$55 &       50 &   0.130 & 34 &   29 &  0.010 &       40 &    7.520 \\
\texttt{gen400\_p0.9\_65} &  400 &   71820 & 0.900 &   $\geq$65 &       54 &   0.030 & 34 &   32 &  0.010 &       40 &    7.520 \\
\texttt{gen400\_p0.9\_75} &  400 &   71820 & 0.900 &   $\geq$75 & {\bf 75} &   0.120 & 36 &   37 &  0.010 &       52 &    7.530 \\
\texttt{hamming8-4}       &  256 &   20864 & 0.639 &         16 &       14 &   0.070 & 10 &   16 &  0.000 &       16 &    3.200 \\
\texttt{keller4}          &  171 &    9435 & 0.649 &         11 & {\bf 11} &   0.000 &  8 &    8 &  0.000 & {\bf 11} &    1.340 \\
\texttt{keller5}          &  776 &  225990 & 0.752 &   $\geq$27 &       23 &   4.570 & 17 &   15 &  0.030 &       20 &   41.880 \\
\texttt{p\_hat300-1}      &  300 &   10933 & 0.244 &          8 &  {\bf 8} &   0.350 &  6 &    7 &  0.000 &  {\bf 8} &    4.650 \\
\texttt{p\_hat300-2}      &  300 &   21928 & 0.489 &         25 & {\bf 25} &   0.150 & 16 &   22 &  0.000 &       22 &    4.980 \\
\texttt{p\_hat300-3}      &  300 &   33390 & 0.744 &         36 & {\bf 36} &   3.160 & 19 &   31 &  0.000 &       30 &    4.320 \\
\texttt{p\_hat700-1}      &  700 &   60999 & 0.249 &         11 & {\bf 11} &   2.330 &  7 &    7 &  0.030 &       10 &   38.060 \\
\texttt{p\_hat700-2}      &  700 &  121728 & 0.498 &         44 & {\bf 44} &   7.690 & 24 &   40 &  0.030 &       35 &   39.110 \\
\texttt{p\_hat700-3}      &  700 &  183010 & 0.748 &   $\geq$62 & {\bf 62} &  13.570 & 31 &   58 &  0.030 &       47 &   37.940 \\
\texttt{p\_hat1500-1}     & 1500 &  284923 & 0.253 &         12 & {\bf 12} &   0.780 &  7 &    9 &  0.180 &       11 &  221.800 \\
\texttt{p\_hat1500-2}     & 1500 &  568960 & 0.506 &   $\geq$65 & {\bf 65} &  18.130 & 30 &   61 &  0.180 &      48 &  224.230 \\
\hline
\end{tabular}
\end{center}
\caption{Results for the DIMACS benchmarks (Second challenge
set)}\label{tabella1}
\end{small}
\end{table}
\end{landscape}

\begin{landscape}
\begin{table}
\begin{small}
\begin{center}
\begin{tabular}{l|cccc|cc|c|cc|cc}
\hline
&&&&&\multicolumn{2}{|c|}{$\cal{C}$} & $\cal{G}$ & \multicolumn{2}{|c|}{$\cal{D}$}& \multicolumn{2}{|c}{$\cal{MC}$}\\
\hline
DIMACS benchmarks     & $n$ & $m$ & $d$ &$\o(G)$&     $k$ & Time($s$)    &          $k$ &          $k$ &  Time($s$)          &          $k$ &  Time($s$)    \\
\hline\hline
\texttt{brock200\_1}    &  200 &  14834 & 0.745 &       21 &  {\bf 21} & 1.800 & 16 &        16 & 0.000 &       17 &  1.970 \\
\texttt{brock200\_3}    &  200 &  12048 & 0.605 &       15 &        14 & 0.360 & 10 &        11 & 0.000 &       13 &  2.240 \\
\texttt{brock400\_1}    &  400 &  59723 & 0.748 &       27 &        25 & 0.400 & 16 &        18 & 0.010 &       21 &  9.150 \\
\texttt{brock400\_3}    &  400 &  59681 & 0.748 &       31 &        25 & 0.400 & 16 &        17 & 0.000 &       20 &  9.200 \\
\texttt{brock800\_1}    &  800 & 207505 & 0.649 &       23 &        21 & 1.390 & 15 &        14 & 0.040 &       17 & 55.970 \\
\texttt{brock800\_3}    &  800 & 207333 & 0.649 &       25 &        22 & 0.880 & 14 &        14 & 0.040 &       18 & 55.920 \\
\texttt{c-fat200-1}     &  200 &   1534 & 0.077 &       12 &  {\bf 12} & 0.000 &  8 &  {\bf 12} & 0.000 & {\bf 12} &  1.740 \\
\texttt{c-fat200-2}     &  200 &   3235 & 0.163 &       24 &  {\bf 24} & 0.010 & 14 &  {\bf 24} & 0.000 &       23 &  1.750 \\
\texttt{c-fat200-5}     &  200 &   8473 & 0.426 &       58 &  {\bf 58} & 0.010 & 32 &  {\bf 58} & 0.000 &       48 &  1.790 \\
\texttt{c-fat500-10}    &  500 &  46627 & 0.374 &      126 & {\bf 126} & 0.020 & 66 & {\bf 126} & 0.010 &       96 & 14.900 \\
\texttt{c-fat500-1}     &  500 &   4459 & 0.036 &       14 &  {\bf 14} & 0.000 &  9 &  {\bf 14} & 0.010 &       12 & 14.760 \\
\texttt{c-fat500-2}     &  500 &   9139 & 0.073 &       26 &  {\bf 26} & 0.020 & 15 &  {\bf 26} & 0.010 &       22 & 14.950 \\
\texttt{c-fat500-5}     &  500 &  23191 & 0.186 &       64 &  {\bf 64} & 0.020 & 34 &  {\bf 64} & 0.010 &       51 & 14.660 \\
\texttt{hamming6-2}     &   64 &   1824 & 0.905 &       32 &  {\bf 32} & 0.020 & 18 &  {\bf 32} & 0.000 &       29 &  0.160 \\
\texttt{hamming6-4}     &   64 &    704 & 0.349 &        4 &   {\bf 4} & 0.000 &  3 &   {\bf 4} & 0.000 &  {\bf 4} &  0.210 \\
\texttt{hamming8-2}     &  256 &  31616 & 0.969 &      128 & {\bf 128} & 0.020 & 58 & {\bf 128} & 0.000 &       97 &  2.270 \\
\texttt{johnson8-2-4}   &   28 &    210 & 0.556 &        4 &   {\bf 4} & 0.010 &  3 &   {\bf 4} & 0.000 &  {\bf 4} &  0.050 \\
\texttt{johnson8-4-4}   &   70 &   1855 & 0.768 &       14 &  {\bf 14} & 0.010 &  9 &         8 & 0.000 & {\bf 14} &  0.220 \\
\texttt{johnson16-2-4}  &  120 &   5460 & 0.765 &        8 &   {\bf 8} & 0.020 &  7 &   {\bf 8} & 0.000 &  {\bf 8} &  0.620 \\
\texttt{johnson32-2-4}  &  496 & 107880 & 0.879 &       16 &  {\bf 16} & 0.890 & 15 &  {\bf 16} & 0.010 & {\bf 16} & 14.350 \\
\texttt{MANN\_a9}       &   45 &    918 & 0.927 &       16 &  {\bf 16} & 0.000 & 12 &        12 & 0.000 & {\bf 16} &  0.090 \\
\texttt{p\_hat500-1}    &  500 &  31569 & 0.253 &        9 &   {\bf 9} & 0.220 &  7 &         7 & 0.010 &  {\bf 9} & 18.170 \\
\texttt{p\_hat500-2}    &  500 &  62946 & 0.505 &       36 &  {\bf 36} & 0.120 & 18 &        32 & 0.010 &       29 & 18.940 \\
\texttt{p\_hat500-3}    &  500 &  93800 & 0.752 & $\geq$50 &  {\bf 50} & 2.280 & 26 &        46 & 0.010 &       39 & 17.080 \\
\texttt{p\_hat1000-1}   & 1000 & 122253 & 0.245 & $\geq$10 &  {\bf 10} & 2.280 &  7 &         8 & 0.080 &        9 & 92.480 \\
\texttt{p\_hat1000-2}   & 1000 & 244799 & 0.490 & $\geq$46 &  {\bf 46} & 0.270 & 23 &        42 & 0.070 &       36 & 94.620 \\
\texttt{p\_hat1000-3}   & 1000 & 371746 & 0.744 & $\geq$68 &  {\bf 68} & 1.950 & 33 &        55 & 0.070 &       49 & 89.710 \\
\texttt{san200\_0.7\_1} &  200 &  13930 & 0.700 &       30 &  {\bf 30} & 0.010 & 16 &        15 & 0.000 &       16 &  2.070 \\
\texttt{san200\_0.7\_2} &  200 &  13930 & 0.700 &       18 &        15 & 0.070 & 13 &        12 & 0.000 &       13 &  2.060 \\
\texttt{san200\_0.9\_1} &  200 &  17910 & 0.900 &       70 &        62 & 0.020 & 39 &        45 & 0.000 &       45 &  1.600 \\
\texttt{san200\_0.9\_2} &  200 &  17910 & 0.900 &       60 &  {\bf 60} & 0.020 & 31 &        35 & 0.000 &       45 &  1.610 \\
\texttt{san200\_0.9\_3} &  200 &  17910 & 0.900 &       44 &        42 & 0.010 & 26 &        24 & 0.000 &       30 &  1.600 \\
\texttt{san400\_0.9\_1} &  400 &  71820 & 0.900 &      100 &        96 & 0.020 & 48 &        50 & 0.000 &       51 &  7.360 \\
\texttt{sanr200\_0.7}   &  200 &  13868 & 0.697 &       18 &  {\bf 18} & 0.080 & 12 &        16 & 0.000 &       16 &  2.070 \\
\texttt{sanr200\_0.9}   &  200 &  17863 & 0.898 &       42 &  {\bf 42} & 0.150 & 27 &        36 & 0.000 &       34 &  1.610 \\
\texttt{sanr400\_0.5}   &  400 &  39984 & 0.501 &       13 &  {\bf 13} & 0.910 &  9 &         8 & 0.000 &       11 & 10.750 \\
\texttt{sanr400\_0.7}   &  400 &  55869 & 0.700 &       21 &  {\bf 21} & 0.150 & 14 &        16 & 0.000 &       17 &  9.550 \\
\hline\end{tabular}
\end{center}
\caption{Results for the DIMACS benchmarks
(continue)}\label{tabella2}
\end{small}
\end{table}
\end{landscape}

\begin{table}[h]
\begin{small}
\begin{center}
\begin{tabular}{l|ccc|cc|c|cc|cc}
\hline
&&&&\multicolumn{2}{|c|}{$\cal{C}$} & $\cal{G}$ & \multicolumn{2}{|c|}{$\cal{D}$}& \multicolumn{2}{|c}{$\cal{MC}$}\\
\hline
Random graph        &   $n$ &      $m$ &  $d$  &$k$ & Time($s$) &   $k$ & $k$ & Time($s$) & $k$ &  Time($s$)    \\
\hline\hline
\texttt{tbb128.5}   &   128 &     4061 & 0.500 & 11 &   0.010 &  7 &  8 &  0.000 & 11 &      7.520 \\
\texttt{tbb256.5}   &   256 &    16310 & 0.500 & 12 &   0.510 &  9 &  9 &  0.000 & 11 &     30.750 \\
\texttt{tbb512.5}   &   512 &    65457 & 0.500 & 13 &   1.150 &  9 & 10 &  0.000 & 12 &    131.980 \\
\texttt{tbb1024.5}  &  1024 &   262084 & 0.500 & 15 &   2.170 & 10 &  9 &  0.060 & 13 &    589.350 \\
\texttt{tbb2048.5}  &  2048 &  1048289 & 0.500 & 16 &   9.280 & 11 & 10 &  0.222 & 14 &   2564.550 \\
\texttt{tbb4096.5}  &  4096 &  4192863 & 0.500 & 17 &  73.840 & 12 & 11 &  1.870 & 15 &  11061.950 \\
\texttt{tbb8192.5}  &  8192 & 16778527 & 0.500 & 19 & 311.950 & 13 & 11 &  8.130 & 16 &  50238.440 \\
\texttt{tbb16384.5} & 16384 & 67106538 & 0.500 & 19 & 170.470 & 14 & 11 & 33.850 & 17 & 216040.910 \\
\hline
\end{tabular}
\end{center}
\caption{Results for random graph with density
$d=0.5$}\label{tabella3}
\end{small}
\end{table}

\begin{table}[h]
\begin{small}
\begin{center}
\begin{tabular}{l|ccc|cc|c|cc|cc}
\hline
&&&&\multicolumn{2}{|c|}{$\cal{C}$} & $\cal{G}$ & \multicolumn{2}{|c|}{$\cal{D}$}& \multicolumn{2}{|c}{$\cal{MC}$}\\
\hline
Random graphs       &   $n$ &      $m$ &  $d$  &$k$ & Time($s$) &   $k$ & $k$ & Time($s$) & $k$ &  Time($s$)    \\
\hline\hline
\texttt{tbb128.9}   &   128 &      7315 & 0.900 & 34 &   0.080 & 24 & 29 &  0.000 &  30 &      5.860 \\
\texttt{tbb256.9}   &   256 &     29392 & 0.900 & 44 &   1.310 & 28 & 37 &  0.000 &  36 &     23.390 \\
\texttt{tbb512.9}   &   512 &    117794 & 0.900 & 56 &   3.190 & 37 & 46 &  0.010 &  42 &     98.940 \\
\texttt{tbb1024.9}  &  1024 &    471440 & 0.900 & 67 &   9.720 & 42 & 49 &  0.060 &  48 &    445.620 \\
\texttt{tbb2048.9}  &  2048 &   1886256 & 0.900 & 76 &  35.740 & 49 & 55 &  0.222 &  54 &   1958.010 \\
\texttt{tbb4096.9}  &  4096 &   7548970 & 0.900 & 84 &  41.780 & 56 & 57 &  1.860 &  59 &   8307.030 \\
\texttt{tbb8192.9}  &  8192 &  30198965 & 0.900 & 90 & 182.350 & 61 & 63 &  8.120 &  66 &  35811.920 \\
\hline
\end{tabular}
\end{center}
\caption{Results for random graph with density $d=0.9$}\label{tabella4}
\end{small}
\end{table}

\begin{table}[h]
\begin{small}
\begin{center}
\begin{tabular}{l|cc|cc}
\hline
&\multicolumn{2}{|c|}{$\cal{C}$} & \multicolumn{2}{|c}{Cliquer}\\
\hline
Random graphs      &$k$ & Time($s$) &  $k$ &  Time($s$)    \\
\hline\hline
\texttt{tbb128.5}  & 11 & 0.010 & 11 &    0.000\\
\texttt{tbb256.5}  & 12 & 0.510 & 12 &    0.130\\
\texttt{tbb512.5}  & 13 & 1.150 & 14 &   10.540\\
\texttt{tbb1024.5} & 15 & 2.170 & 15 & 1769.910\\
\hline
\hline
\texttt{tbb128.9}  & 34 &  0.080 & 34 & 61.570\\
\hline
\end{tabular}
\end{center}
\caption{Comparison between $\cal{C}$ and Cliquer on some random
instances}\label{tabella5}
\end{small}
\end{table}

\vglue15.truecm


\begin{thebibliography}{99}

\bibitem{bollobas}
{\sc B. Bollobas}
\newblock{\em Random graph},
\newblock 2nd ed.,Cambridge University Press, 2001.

\bibitem{CF}
{\sc M. Caramia, G. Felici}
\newblock {\em Mining relevant information on the Web: a clique based approach},
\newblock International Journal of Production Research,
special issue on Data Mining, accepted 2006.

\bibitem{dankel}
{\sc P. Dankelmanna, G. S. Domkeb, W. Goddardc, P. Groblerd, J. H.
Hattinghb, H. C. Swarta},
\newblock {\em Maximum sizes of graphs with given domination parameters},
\newblock Discrete Mathematics 281 (2004), 137-148.

\bibitem{GJ}
{\sc M. R. Garey, D.S. Johnson},
\newblock {\em Computer and Intractability: A guide to the theory of
NP-completeness},
\newblock Freeman, New York, 1976.

\bibitem{ISS}
{\sc A. Iovanella, B. Scoppola, E. Scoppola},
\newblock {\em },
\newblock in preparation.

\bibitem{jerrum}
{\sc M. Jerrum},
\newblock {\em Large cliques elude the metropolis process},
\newblock Random Structures and Algorithms, 3, 4 (1992), 347--359.

\bibitem{mixing}
{\sc M. Jerrum and A. Sinclair},
\newblock {\em The Markov chain Monte Carlo method: an approach to approximate
counting and integration},
\newblock in "Approximation Algorithms for NP-hard Problems," D.S.Hochbaum ed., PWS Publishing, Boston, 1996.

\bibitem{johnson}
{\sc D. S. Johnson, M. Trick (eds.)},
\newblock {\em Cliques, coloring and satisfiability: Second DIMACS
implementation challenge},
\newblock DIMACS Series in Discrete Mathematics and Theoretical
Computer Science, vol. 26, American Mathematical Society, Providence, RI, 1996.

\bibitem{KGV}
{\sc S. Kirkpatric, C. D. Gellatt, M. P. Vecchi},
\newblock {\em Optimization by simulated annealing},
\newblock Science 220, 671-680.

\bibitem{MP}
{\sc M. M\'ezard, G. Parisi},
\newblock {\em The cavity method at zero temperature},
\newblock  J. Stat. Phys 111 (2003) 1.

\bibitem{MPV}
{\sc M. M\'ezard, G. Parisi, M. Virasoro},
\newblock {\em Spin Glass Theory and  Beyond},
\newblock World Scientific, (1987).

\bibitem{MPZ}
{\sc M. M\'ezard, G. Parisi, R. Zecchina},
\newblock {\em Analytic and algorithmic solution of random
  satisfiability
problems},
\newblock Science 297, 812-815 (2002).

\bibitem{MZ}
{\sc M. M\'ezard, R. Zecchina},
\newblock {\em The random K-satisfiability problem: from an analytic
solution to an efficient algorithm},
\newblock Phys.Rev. E 66, 056126-1/056126-27 (2002).

\bibitem{NO}
{\sc S. Niskanen, P. R. J. \"Osterg\"ard},
\newblock {\em Cliquer User's Guide, Version 1.0},
\newblock Communication Laboratory, Helsinki University of Technology,
Espoo, Finland, Tech. Rep. T48, 2003.

\end{thebibliography}
\end{document}